\documentclass[aps,pra,twocolumn,superscriptaddress,nofootinbib,amsmath,amssym,floatfix]{revtex4-2}

\usepackage{enumitem}                  
\usepackage{float}                     
\usepackage{graphicx}                  
\usepackage{graphics}                  
\usepackage{amsmath}                   
\usepackage{commath}                   
\usepackage{bm}
\usepackage{booktabs}                  
\usepackage{multirow}
\usepackage{datetime}                  
\usepackage{hyperref}
\usepackage[export]{adjustbox}         
\usepackage{xcolor}                    
\let\oldhat\hat
\renewcommand{\hat}[1]{\oldhat{\mathbf{#1}}}
\usepackage[normalem]{ulem} 

\begin{document}

\title{Breakdown of light transport models in photonic scattering slabs with strong absorption and anisotropy}

\author{Ozan Akdemir}
\email{o.akdemir@utwente.nl}
\affiliation{Complex Photonic Systems (COPS), MESA+ Institute for Nanotechnology, 
University of Twente, P.O. Box 217, 7500 AE Enschede, The Netherlands}

\author{Ad Lagendijk}
\affiliation{Complex Photonic Systems (COPS), MESA+ Institute for Nanotechnology, 
University of Twente, P.O. Box 217, 7500 AE Enschede, The Netherlands}

\author{Willem L. Vos}
\email{w.l.vos@utwente.nl}
\affiliation{Complex Photonic Systems (COPS), MESA+ Institute for Nanotechnology, 
University of Twente, P.O. Box 217, 7500 AE Enschede, The Netherlands}

\date{30 January 2022} 

\begin{abstract}

The radiative transfer equation (RTE) models the transport of light inside photonic scattering samples such as paint, foam and tissue. 
Analytic approximations to solve the RTE fail for samples with strong absorption and dominant anisotropic scattering and predict unphysical negative energy densities and the diffuse flux in the wrong direction. 
Here we fully characterize the unphysical regions of three popular approximations to the RTE for a slab, namely the $P_1$ approximation (or diffusion approximation), the $P_3$ approximation, and a popular modification to $P_3$ that corrects the forward scattering in the approximation. 
We find that the delta function correction to $P_3$ eliminates the unphysical range in the forward scattering.
In addition, we compare the predictions of these analytical methods to exact Monte Carlo simulations for the \textit{physical} and \textit{unphysical} regions.
We present maps of relative errors for the albedo and the anisotropy of the scatterers for a realistic index contrast typical of a polymer slab in air and optical thickness. 
The relative error maps provide a guideline for the accuracy of the analytical methods to interpret experiments on light transport in photonic scattering slabs. 
Our results show that the $P_1$ approximation is significantly inaccurate to extract transport parameters unless the sample scatters purely isotropic and elastic. 
The $P_3$ approximation exceeds $P_1$ in terms of accuracy in its physical range for moderate absorption, and the $P_3$ with the delta function correction is the most accurate approximation considered here for the forward direction. 

\end{abstract}

\maketitle

\section{Introduction}~
\label{section:Introduction}
\begin{figure}[htb!]
\includegraphics[width=1\columnwidth]{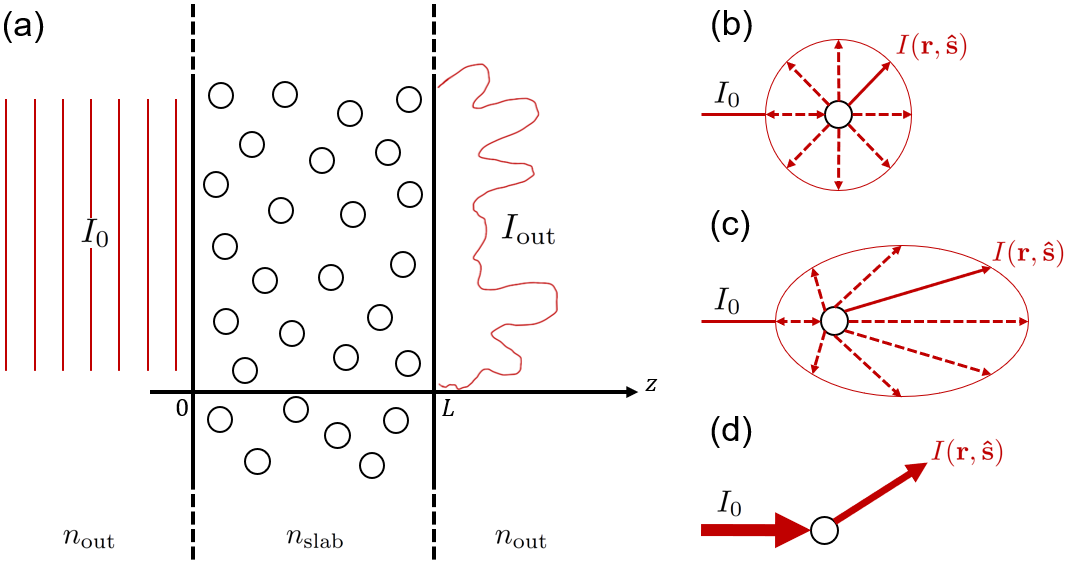}
\caption{(a) Incident plane waves with intensity $I_0$ are scattered by scatterers (spheres) inside a slab and leave as scattered waves (speckle) with intensity $I_{\textrm{out}}$. 
The refractive indices of the slab and of the medium outside are $n_{\textrm{slab}}$ and $n_{\textrm{out}}$, respectively. 
(b) An isotropic scatterer that scatters in all directions with equal probability, corresponding to a constant phase function. 
(c) An anisotropic scatterer that scatters more in the forward direction. 
(d) An absorbing scatterer, where the loss of intensity is depicted as thinner arrows after the scattering event.
In (b,c) the solid arrows are incident and scattered ($\hat{s}$) directions, the dashed arrows are other possible scattering directions, and the arrow lengths indicate the probability to scatter into that direction.
The centers of the scattering spheres in (b,c,d) are at $\mathbf{r}$.
}
\label{fig:Illustration}
\end{figure}

In photonic scattering media, such as paint, foam and tissue, the refractive index varies spatially causing incident waves to be scattered and absorbed~\cite{Ishimaru1978book,vanRossum1999RMP,Akkermans2007Book,Wiersma2013NP,Ghulinyan2015BookChp,Carminati2021Book}. 
Understanding the transport of light in such scattering media is crucial for many application areas, such as atmospheric and climate sciences~\cite{Fournier1994OOXII,Fell2001JQSRT,Stramski2004PO,Nousiainen2015Book}, 
oceanography~\cite{Funk1973AO,McCormick2006Book}, biophysics~\cite{Star1989BookChp,Jacques1990JQE,Cheong1990JQE,Kienle2006PRL}, powder technology
\cite{Sekulic1996AC,Burger1997AS}, and solid-state lighting~\cite{Schubert2006Book, Krames2007JDT, Bechtel2008Conference, Meretska2019ACSPhot}. 
The transport theory describes the propagation of wave in scattering media, notably in a widely-used realistic situation like a slab in three dimensions (3D), as shown in Fig.~\ref{fig:Illustration}(a). 
The theory describes the transfer of intensity and neglects interference effects including diffraction~\cite{Ishimaru1978book}. 
For all practical purposes the transport theory is rigorous and only in exceptional cases such as the situations of very strong elastic scattering and Anderson localization of light where interference predominates, the transport shows features beyond the predictions of transport theory~\cite{vanRossum1999RMP}. 

The basic differential equation used in transport theory is the radiative transfer equation (RTE), which is equivalent to Boltzmann’s equation used in the kinetic theory of gases and neutron transport~\cite{Ishimaru1978book,Chandrasekhar1960Book,Tait1964Book}.
The most fundamental quantity is the specific intensity $I(\mathbf{r,\hat{s}})$ that describes the average power flux density at position ${\bm r}$ in a given direction $\mathbf{\hat{s}}$ within a unit solid angle and a unit frequency band~\cite{Ishimaru1978book}.
Due to its dependency on both the position $\mathbf{r}$ and the direction $\hat{s}$, it is challenging to solve $I(\mathbf{r,\hat{s}})$ directly.

The most popular method to solve the RTE for light is the Monte Carlo simulation of light transport~\cite{Prahl1989BookChp,Mujumdar2010JN,Jacques2011BookChp,Atif2011OS,Uppu2013PRA,Leino2019OSAC,Jonsson2020OE,Cooper2021C,Krieger2021AA}, a statistical method that converges to the exact solution of the RTE. 
To obtain a high accuracy, however, this method comes with the cost of extremely long computation times~\cite{Shonkwiler2009Book}, high computational power requirements with concomitant high energy consumption.

The complexity of transport theory and the tedious resource-consuming Monte Carlo simulations have stimulated the development of analytical approximations to the RTE~\cite{Joseph1976JAS,Star1988PMB,Keijzer1988AO,Dickey1998PMB,Liemert2011PRA,Liemert2021JOSAA}.
These analytical approximations are sustainable alternatives to Monte Carlo simulations, since computations consume much less resources.
In addition, despite impressive advances made by graphics processing unit (GPU) based Monte Carlo simulations in terms of speed~\cite{Leino2019OSAC,Jonsson2020OE}, analytical methods are significantly faster. 
Moreover, in certain configurations, such as the slab geometry, the results of these analytical approximations match the accuracy of simulations. 

A widely used analytical approximation to the RTE is the $P_N$ approximation (See Appendix \ref{appendix:PNapproximation} for full solution for a slab geometry), where the dependence on both variables is separated~\cite{Tait1964Book} by expanding the specific intensity $I(\mathbf{r,\hat{s}})$ in products of complete sets on the domains of $\bm{r} $ and $\bm {\hat s}$
%
\begin{equation}{\label{eqn:specIntExp}}
 I(\mathbf{r,\hat{s}}) = \sum_{l = 0}^{N} \sum_{m = -l}^{l}
 \psi_{l}^{m}(\mathbf{r}) Y_{l}^{m} (\mathbf{\hat{s}}).
\end{equation}
%
Here, $\psi_{l}^{m}(\mathbf{r})$ are the spatial components, $Y_{l}^{m} (\mathbf{\hat{s}})$ are the Laplace spherical harmonics~\cite{Jackson1998Book,Arfken2005Book}, and $N$ is the order of the approximation that determines the number of terms in Eq.~(\ref{eqn:specIntExp}). 
The analytical $P_N$ approximations are mostly used for simple sample geometries such as a slab and a sphere, and their accuracies depend notably on (\textit{i}) the order $N$ of the approximation and (\textit{ii})  on the optical properties of the medium. 
As $N$ approaches infinity, the $P_N$ approximation yields exact solutions.
In realistic cases, such as a 3D slab, the order is rarely higher than $N=3$ as the mathematical complexity increases rapidly with increasing $N$ and becomes computationally demanding.~\footnote{Odd positive integers are chosen for $N$ since odd order approximations are known to be more accurate than even orders, as in the latter the angular integrands are discontinuous~\cite{Case1967Book}.} 
In this work, we thoroughly validate the $P_1$ and $P_3$ approximations for a slab geometry~\cite{ExtentionNote}.
%

The first-order analytical approximation $P_1$ to transport theory is the diffusion theory~\cite{Ishimaru1978book}, which is widely used to extract transport parameters from opaque media with isotropic scatterers (see Fig.~\ref{fig:Illustration}(b)) and with negligible absorption.
For a slab shown in Fig.~\ref{fig:Illustration}(a), this opaque configuration amounts to the thickness $L$ being much larger than the transport mean free path ${\ell}_{\textrm{tr}}$ and much smaller than the absorption mean free path ${\ell}_{\textrm{abs}}$:
%
\begin{equation}\label{GeneralDifLimits}
{\ell}_{\textrm{tr}} \ll L \ll {\ell}_{\textrm{abs}}\text{.}
\end{equation}
%
If the scattering is dominantly in the forward direction, ${\ell}_{\textrm{tr}}$ increases (see Fig.~\ref{fig:Illustration}(c)), and ${\ell}_{\textrm{abs}}$ decreases if the scatterers have significant absorption (see Fig.~\ref{fig:Illustration}(d)).
Meretska \textit{et al.}~\cite{Meretska2019arXiv} defined a physically more informative validity range using the three-parameter space ($a$,$g$,$b$) spanned by the albedo $a$, the anisotropy $g$, and the optical thickness $b$. 
A practically relevant parameter space also requires consideration of the internal reflection at the slab boundaries~\cite{Lagendijk1989PLA,Zhu1991PRA}, so we add as a 4\textsuperscript{th} parameter the refractive index contrast $\Delta n^2$
%
\begin{equation}\label{eqn:Dn}
\Delta n^2 \equiv \frac{n_{\textrm{slab}}^2 - n_{\textrm{out}}^2}{2 n_{\textrm{slab}}^2}.
\end{equation}
Here, $n_{\textrm{slab}}$ and $n_{\textrm{out}}$ are the refractive index of the medium inside the slab that surrounds the scatterers and the index of the medium outside the slab (typically free space), respectively. 
Once the ($a$,$g$,$b$,$\Delta n^2$) parameter set is known, the solution of $P_N$ approximation is fully determined.

\begin{figure}[htb!]
\centering
\includegraphics[width = 1\columnwidth]{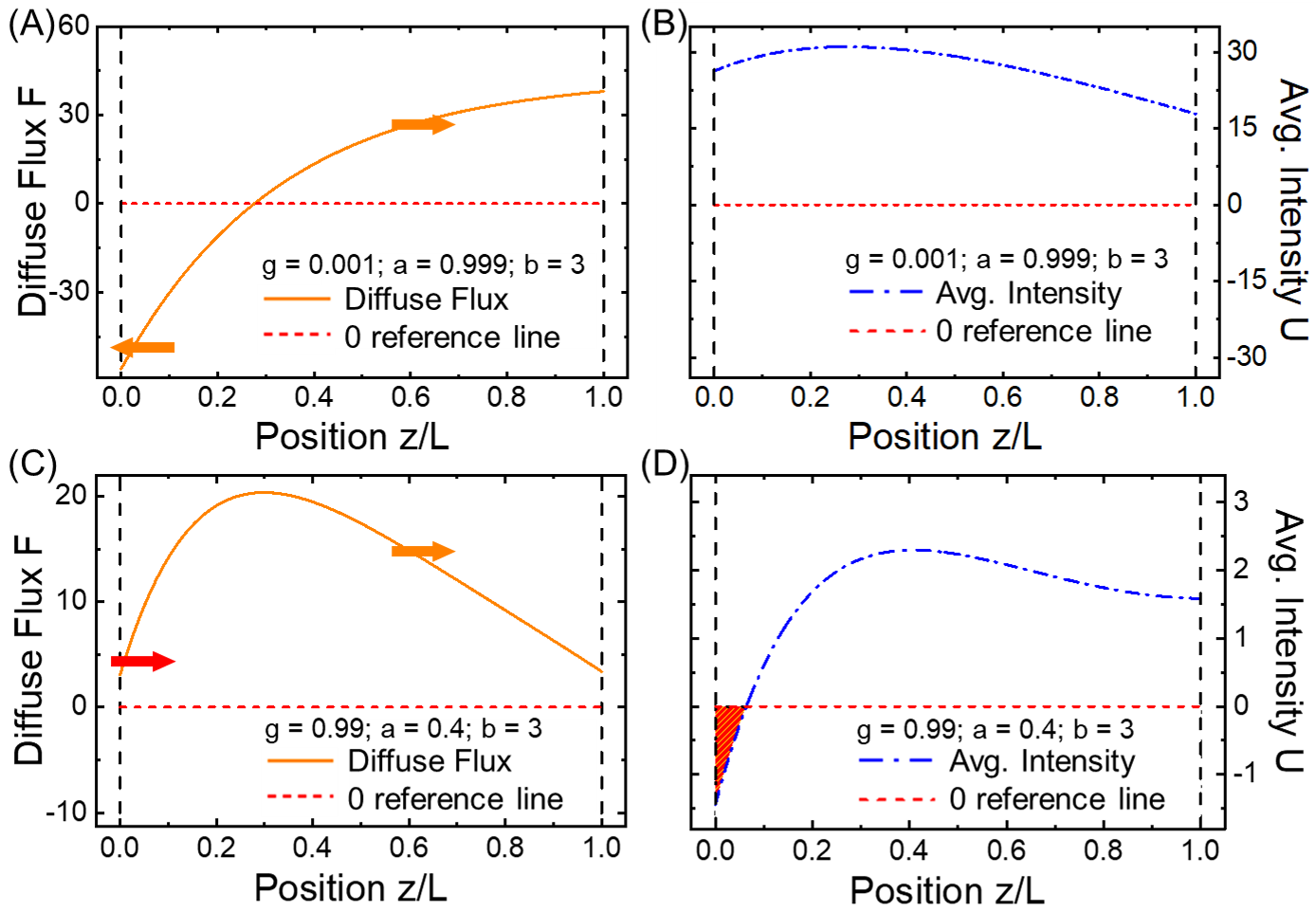}
\caption{
Examples of physical and unphysical results. 
(A) Diffuse flux $F$ and (B) average intensity $U$ computed with the $P_1$ or diffusion approximation for isotropic ($g = 0.001$) scattering with little absorption ($a = 0.999$) are physically sensible. 
(C) Diffuse flux $F$ and (D) average intensity $U$ using $P_1$ for anisotropic scattering ($g = 0.99$) and strong absorption ($a = 0.4$).  
In (C) the red arrow on the left boundary indicates the \textit{unphysical} direction of $F$ and  in (D) the \textit{unphysical} negative average intensity $U$ is highlighted with red hatches. 
Black dashed vertical lines represent the boundaries of the slab with optical thickness $b = 3$ and $\Delta n^2 = 0.245$ typical of a polymer slab in air. 
The direction of $F$ is given by the orange arrows, 
and the red dashed horizontal line indicates the zero level.
}
\label{fig:ExamplesP1}
\end{figure}

In some parts of the ($a$,$g$,$b$,$\Delta n^2$) parameter space, the $P_N$ approximations predict unphysical behavior, such as a negative energy density. 
We call these regions \textit{unphysical ranges}. 
An example of an unphysical result of the $P_1$ approximation is shown in Fig.~\ref{fig:ExamplesP1}(C,D). 
In Fig.~\ref{fig:ExamplesP1}(D) the diffusion theory predicts an unphysical negative energy density. 
The diffuse flux $F$ given in Fig.~\ref{fig:ExamplesP1}(C) is unphysical, since the theory predicts erroneously an {\it incident} diffuse flux $F$, whereas a reflected flux opposite to the incident direction of light is required at the incident boundary (left boundary in Fig.~\ref{fig:ExamplesP1}). 
%

In this paper, we map the unphysical ranges and the relative errors of three popular approximations to the RTE, namely, the $P_1$, the $P_3$, and the $P_3+\delta E(4)$ approximations. 
The $P_3$ approximation is a widely used higher-order approximation to RTE~\cite{Bayazitoglu79AIAAJ, Meretska2019ACSPhot, Meretska2017OE, Liemert2014MP}, especially popular in biophysics~\cite{Star1989BookChp,Star1988PMB,Dickey1998PMB}. 
The $P_3 + \delta E(4)$ approximation uses a modified phase function within the $P_3$ approximation to increase the accuracy of the forward scattering region~\cite{Star1989BookChp,Star1988PMB,Meador1979AO} (see Appendix \ref{appendix:PNdE(4)}). 
The unphysical ranges of each approximation are found by scanning the ($a$,$g$,$b$,$\Delta n^2$) parameter space and looking for the regions where the conditions (\ref{eqn:PNLimits}) are violated. 
In addition, the relative error maps are obtained by comparing the $P_N$ approximations to extensive Monte Carlo simulations. 
Our simulation code is based on the work of Prahl \textit{et al.}~\cite{Prahl1989BookChp}, with the addition of multiple internal reflection similar to a Fabry-Pérot cavity~\cite{Hecht2016Book}, where only the multiply reflected intensities are considered but no interference, see Ref. \cite{BohrenHuffman1998Book}. 

\section{Results \& Discussions}
\subsection{The unphysical ranges}
\begin{figure}[htb!]
    \includegraphics[width=1\columnwidth]{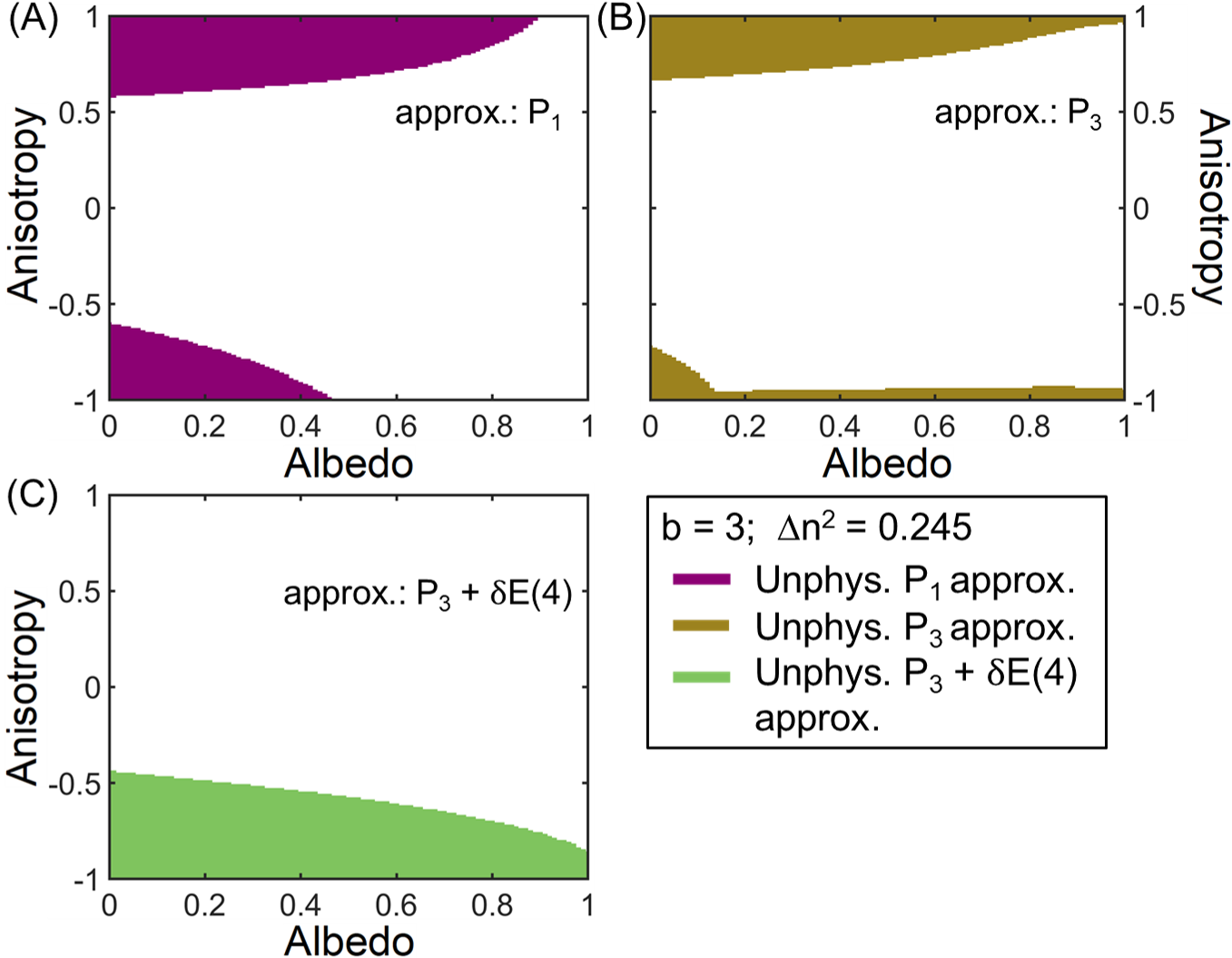}
    \caption{Unphysical ranges of (A) $P_1$ approximation, (B) $P_3$ approximation and (C) $P_3 + \delta E(4)$ approximation.
    Fixed optical thickness $b = 3$ and refractive index contrast $\Delta n^2 = 0.245$
    (\textit{e.g.}, polymer slab in air). 
    In this figure and the following ones, the extreme anisotropy limits are $g = 0.999999$ and $g = -0.999999$, as the exact forward scattering $g = 1$ and exact backscattering $g = -1$ cases are unphysical.
    }
    \label{fig:UnphysicalRanges_FixedOptTDn}
\end{figure}

\begin{figure}[htb!]
    \includegraphics[width=1\columnwidth]{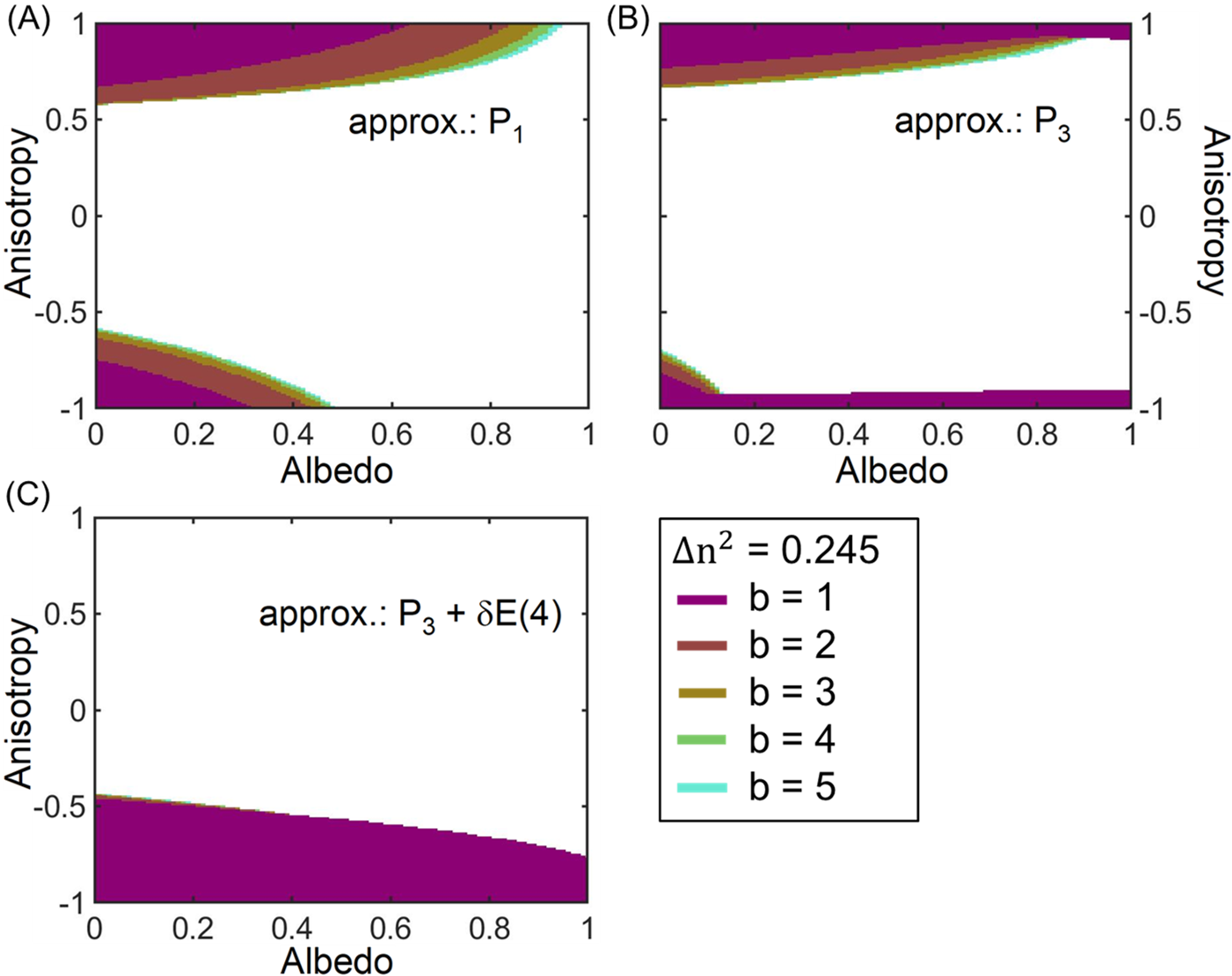}
    \caption{Unphysical ranges of (A) $P_1$ approximation, (B) $P_3$ approximation and (C) $P_3 + \delta E(4)$ approximation, for
    various optical thicknesses and for fixed refractive index contrast $\Delta n^2 = 0.245$
    (\textit{e.g.}, polymer slab in air).
    }
    \label{fig:UnphysicalRanges_VariousOptTFixedDn}
\end{figure}

\begin{figure}[htb!]
    \includegraphics[width=1\columnwidth]{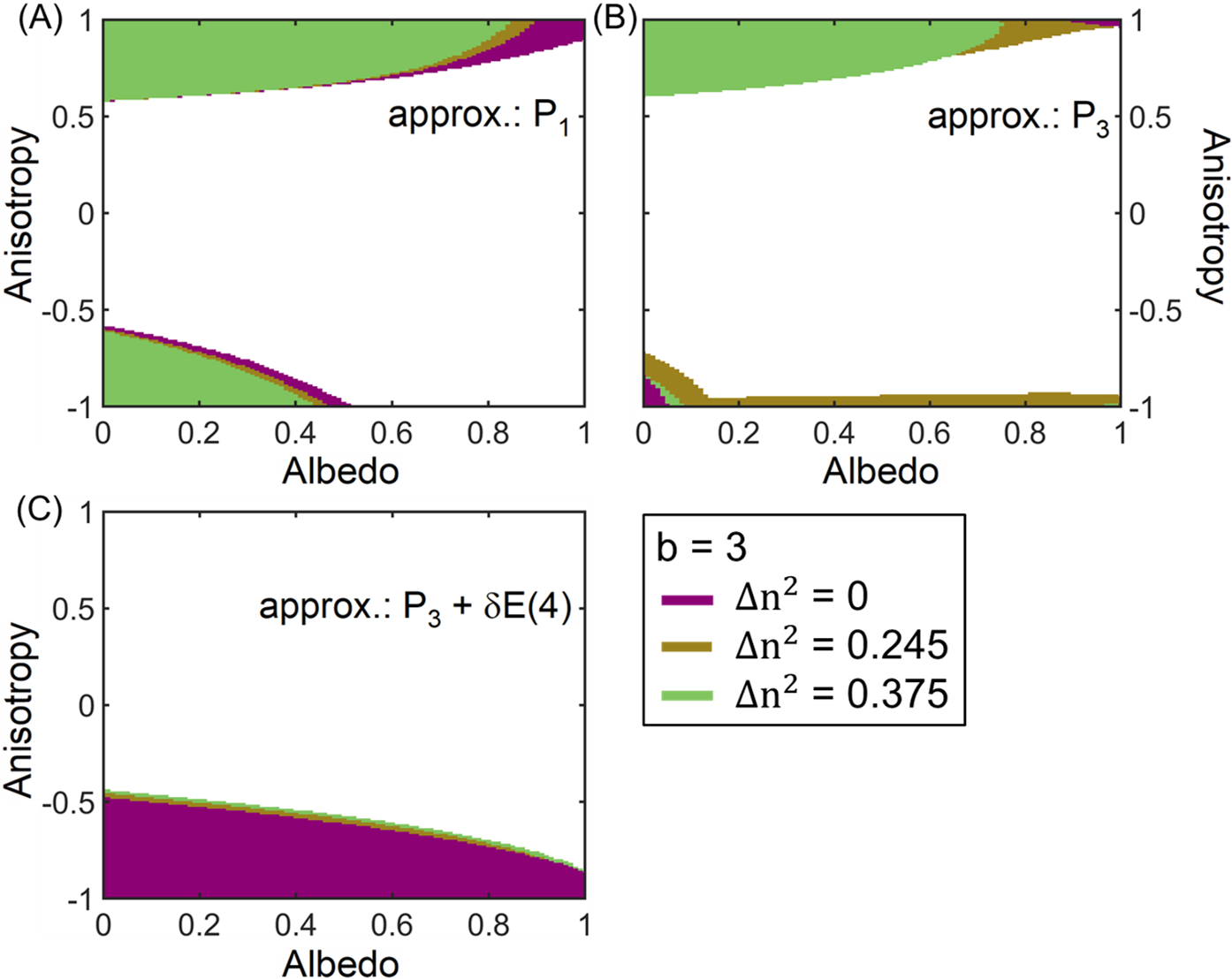}
    \caption{Unphysical ranges of (A) $P_1$ approximation, (B) $P_3$ approximation and (C) $P_3 + \delta E(4)$ approximation, 
    for various refractive index contrasts and for fixed optical thickness $b = 3$.}
    \label{fig:UnphysicalRanges_FixedOptTVariousDn}
\end{figure}

The physical validity conditions of the $P_N$ approximation are expressed mathematically as~\cite{Meretska2019arXiv}:
%
\begin{equation}{\label{eqn:PNLimits}}
\begin{split}
 U(z) &\geq 0 \:\: \forall \:\: z \in [0,L], \\
 F(z) &< 0 \:\: \text{at} \:\: z = 0, \\
 F(z) &> 0 \:\: \text{at} \:\: z = L. 
\end{split}
\end{equation}
%
Here $U(z)$ is the average intensity that is directly proportional to the energy density $u(z)$~\cite{Ishimaru1978book};
%
\begin{equation}\label{eqn:Uvsu}
U(z) \equiv \frac{c}{4\pi} u(z),
\end{equation}
%
where $c$ is the speed of light.
The unphysical ranges are found by checking whether the approximations violate one or more of the conditions given in equation (\ref{eqn:PNLimits}), for each set of parameters($a,g,b,\Delta n^2$). 
The unphysical regions given in Figs.~\ref{fig:UnphysicalRanges_FixedOptTDn}, \ref{fig:UnphysicalRanges_VariousOptTFixedDn} and \ref{fig:UnphysicalRanges_FixedOptTVariousDn} cover all possible albedos and anisotropies in photonic scattering slabs from backscattering to forward scattering. 
For all figures with anisotropy-albedo maps, the limits of anisotropy are $g = -0.999999$ and $g = 0.999999$, and extreme absorption limit is $a = 0.000001$.
The perfect anisotropy ($g = 1$ and $g = -1$) and perfect absorption ($a = 0$) cases are not discussed here, as they are unphysical \cite{Olmos-Trigo2020PRL, Kerker1983JOSA} and thus samples with such properties can not be realized. 
For similar reason~\cite{vanderMolen2006OL}, we do not consider samples with gain.

Fig.~\ref{fig:UnphysicalRanges_FixedOptTDn} shows the unphysical ranges when the optical thickness and refractive index contrast are fixed ($b = 3$; $\Delta n^2 = 0.245$, \textit{e.g.}, a polymer slab in air). 
The $P_3$ approximation, being the higher-order approximation, is generally believed to be an improvement on the $P_1$ approximation \cite{Star1989BookChp}. 
Hence we expected a shrinking of the unphysical range in going from $P_1$ to $P_N$. 
In Fig. \ref{fig:UnphysicalRanges_FixedOptTDn}(b), however, the expected improvement is not observed. 
It is even more remarkable that parts of the physical regions of the $P_1$ approximation are unphysical for the $P_3$. 
As expected, the $P_3+\delta E(4)$ approximation is entirely physical for the forward scattering region. 
In the dominant backscattering range around $g = -0.5$ and below, however, the approximation is primarily unphysical even without absorption, which is also reasonable since the $P_3+\delta E(4)$ approximation is a modification to correct only the forward direction.

%
%

The effects of varying the thickness $b$ and the index contrast $\Delta n^2$ are presented in Fig. \ref{fig:UnphysicalRanges_VariousOptTFixedDn} and Fig. \ref{fig:UnphysicalRanges_FixedOptTVariousDn}, respectively. 
Increasing the optical thickness from $b = 1$ to $b = 5$ (see Fig. \ref{fig:UnphysicalRanges_VariousOptTFixedDn}) has more effect on the unphysical range of the $P_1$ approximation compared to the other methods. 
A clear increase is seen in the unphysical range of the $P_1$ approximation, from strong to weak absorption regions, and some minor expansion towards the isotropic regions with increased thickness $b$. 
For the $P_3$ approximation, increasing $b$ also results in an expansion of the unphysical range, however, the change is smaller than in the $P_1$ approximation. 
In contrast, the unphysical range of the $P_3+\delta E(4)$ approximation remains almost the same for all thicknesses.
%
%
%
%

In Fig. \ref{fig:UnphysicalRanges_FixedOptTVariousDn} we show the unphysical ranges when the index contrast is varied up to $0.375$ at a fixed optical thickness ($b = 3$). 
As $\Delta n^2$ increases, the unphysical range of the $P_1$ approximation shrinks towards the non-absorbing region. 
Interestingly, for the $P_3$ approximation, the $\Delta n^2 = 0$ situation has a very small unphysical range in the extreme limits of the ($a$,$g$) plane.
As the index contrast increases to $\Delta n^2 = 0.245$, the unphysical range increases dramatically. 
However, a further increase to $\Delta n^2 = 0.375$ decreases the unphysical range. 
The unphysical ranges of the $P_3+\delta E(4)$ approximation are again invariant to changes in $\Delta n^2$.

Our investigations of the unphysical ranges show that the $P_1$ approximation to the RTE should not be used for slabs with anisotropic scattering. 
The $P_3$ approximation seems to be especially good for samples with a refractive index matching with the medium outside, as it is physical almost everywhere. 
Nevertheless, detailed investigations (not shown here) of the $P_3$ approximation for $\Delta n^2 = 0$ show that there are significant relative errors in the dominant forward scattering range ($g>0.5$). 
The $P_3 + \delta E(4)$ approximation is found to be physical for the forward scattering region and invariant for the changes in $b$ and $\Delta n^2$. 
It is, however, the worst approximation considered here for the backscattering region.

\subsection{Evaluating relative errors}
\label{subsection:ErrorsAndComparisonExplanation}

In practice, the transport mean free path ${\ell}_{\textrm{tr}}$ and the absorption mean free path ${\ell}_{\textrm{abs}}$ are extracted from total transmission and total reflection experiments using a $P_N$ approximation~\cite{Meretska2017OE}.
The unphysical ranges of a $P_N$ approximation are essential to recognize for which photonic scattering slabs that particular $P_N$ approximation should not be used for extracting these transport parameters. 
However, the bare acknowledgement of the unphysical ranges do not give any information on the accuracy of these methods to interpret observations on samples residing in the physical regions.
Therefore, we extensively compare with Monte Carlo simulations to obtain relative error maps of transport parameters obtained by the analytical approximation methods. 
The optical thickness and refractive index contrast are  chosen to be ($b = 3, \Delta n^2 = 0.245$). \\
%
\begin{figure}[htb!]
    \includegraphics[width=1\columnwidth]{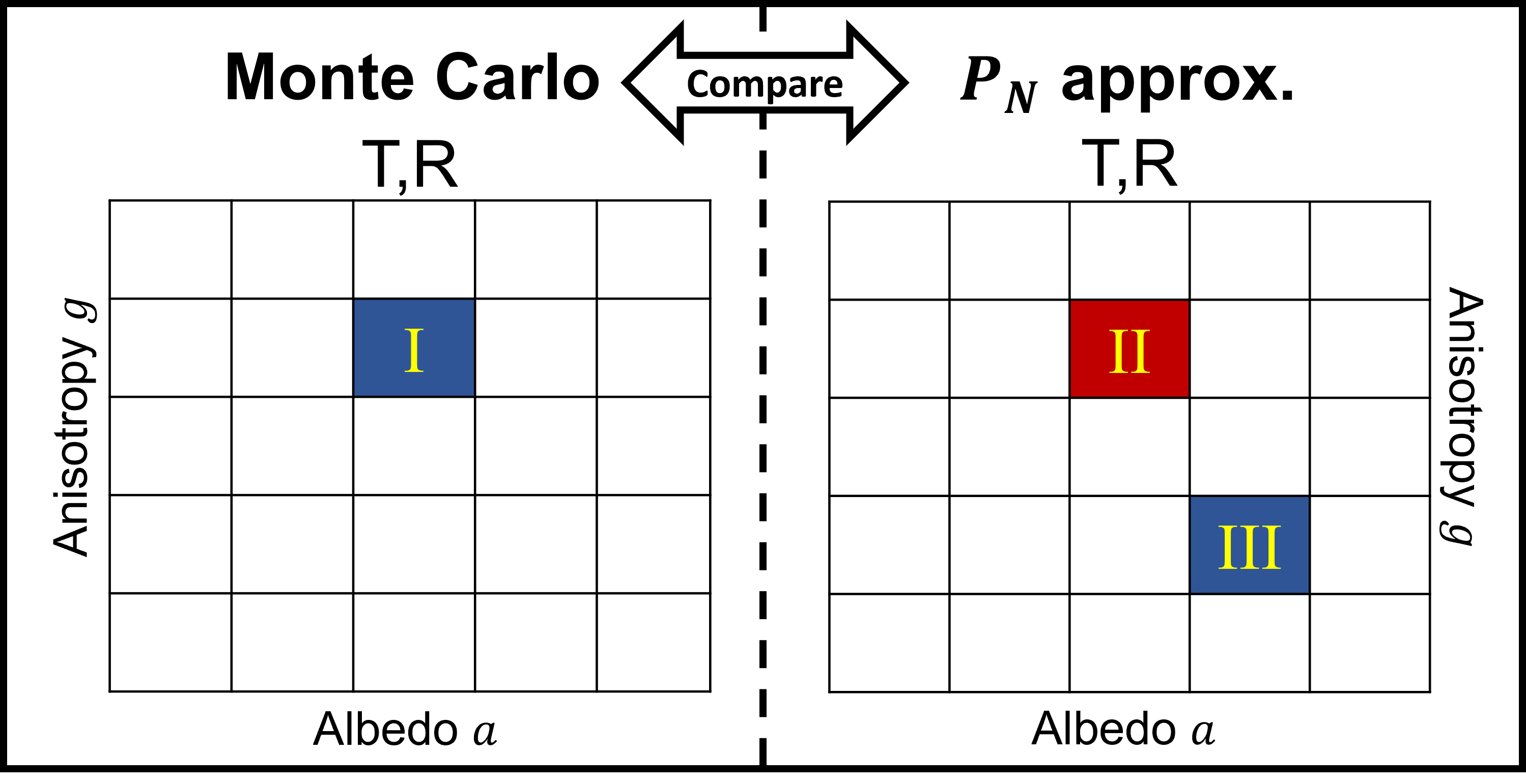}
    \caption{
    Illustration of the comparison process of the analytical methods and Monte Carlo simulations. The albedo $a$ and anisotropy $g$ parameters are used as coordinates for the comparisons. The data on the grids represent the total transmission and total reflection results (T,R). The grid on the left presents the results of the Monte Carlo calculations and the right one presents the results of $P_N$ approximation. The comparison process is explained in detail in subsection \ref{subsection:ErrorsAndComparisonExplanation}.
    }
    \label{fig:ExplanationComparison}
\end{figure}
%
Comparisons are done on the basis of the observation of both the total transmission $T$ and total reflection $R$ that can come from a real experiment or from Monte Carlo simulations. 
The comparison takes four steps: 
\begin{enumerate}
    \item
    We calculate $T$ and $R$ using Monte Carlo simulations, using a fine grid in parameter space~$(a,g)$: $T^{\textrm{MC}}\{a,g\}$ and $R^{\textrm{MC}}\{a,g\}$. 
    We compare these Monte Carlo results to the $P_N$ results obtained for all albedos and anisotropies, using the same fine grid in parameter space~$(a',g')$: $T^{\textrm{P}_N}\{a',g'\}$ and $R^{\textrm{P}_N}\{a',g'\}$. 
    We define the squared relative distance between the two results 
    %
   \begin{eqnarray}{\label{eqn:TandR_DistanceSquared}}
            \delta T(a,g,a',g') &\equiv \frac{   \left[T^{\textrm{MC}}(a,g) -
            T^{\textrm{P}_N}(a',g')\right]^2     }{ | T^{\textrm{MC}}(a,g) |^2  },  \\
            \delta R(a,g,a',g') &\equiv \frac{   \left[R^{\textrm{MC}}(a,g) -
            R^{\textrm{P}_N}(a',g')\right]^2     }{ | R^{\textrm{MC}}(a,g) |^2   }.        
    \end{eqnarray}
    %
    We also define the overall relative distance between the two calculations     
    %
    \begin{equation}{\label{eqn:TandR_Distance}}
               S(a,g,a',g') \equiv  \sqrt {\delta T(a,g,a',g') + \delta R(a,g,a',g')   }.         
    \end{equation}
    %
    %
    \item 
    As a first impression of the errors, the results are compared by calculating relative errors~\cite{Epperson2013Book} for identical grid points $a = a'$ and $g = g'$, defined by        
    %
    \begin{eqnarray}{\label{eqn:TandR_ErrorCalc}}
        \Delta T(a,g) &\equiv \sqrt{ \delta T(a,g,a,g) } \times 100,\\
        \Delta R(a,g) &\equiv \sqrt{ \delta R(a,g,a,g) } \times 100.
    \end{eqnarray}
    %
    An example of a comparison in this step corresponds to comparing the blue square \textrm{I} in the Monte Carlo grid with the red square \textrm{II} in the $P_N$ grid in Fig.~\ref{fig:ExplanationComparison}. 
    The relative error maps for all $(a,g)$ are shown in Fig.~\ref{fig:ErrorMapTT&R}(a-f).
%
\begin{figure*}[htb!]
    \includegraphics[width=1\textwidth]{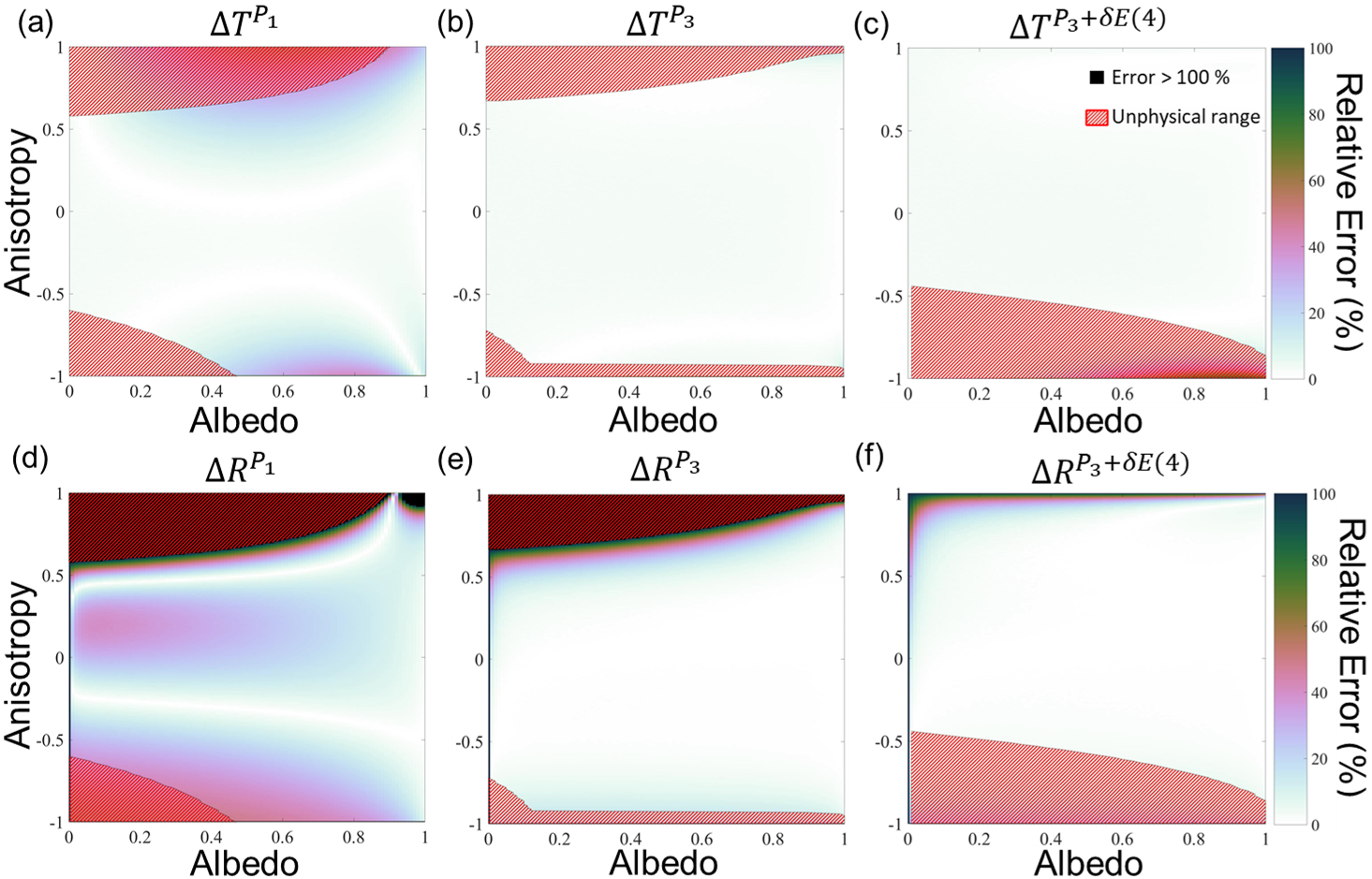}    \caption{Relative error maps of the total transmission and the total reflection results for the $P_N$ approximations.
    (a) and (d) for $P_1$, (b) and (e) for $P_3$, (c) and (f) for  $P_{3}+\delta E(4)$. 
    Relative errors reported here are calculated by comparing these approximations to our Monte Carlo simulations.
    Color map indicates the percentage of errors. Values that are larger than $100\%$ are indicated with black markers. 
    The unphysical ranges of the approximations are added to the graphs as red hatched regions. 
    Optical thickness and refractive index contrast are specifically chosen as $b = 3$ and $\Delta n^2 = 0.245$ for these relative error maps.}
    \label{fig:ErrorMapTT&R}
\end{figure*}
%
    
    \item In a real experiment, one measures the total transmission T and the total reflection R of a sample and uses a numerical or an analytical method to infer the $(a,g)$ parameters that corresponds to the observations. 
    In this work we take T and R from Monte Carlo simulations as they are highly accurate (see Table~\ref{tab:VerificationTable}).
    We call the set obtained from Monte Carlo simulations $(a_\textrm{I},g_\textrm{I})$, from which we infer the true transport lengths, ${\ell}_{\textrm{tr}}$ and ${\ell}_{\textrm{abs}}$. 
    If we interpret the Monte Carlo T and R results with the $P_N$ approximation, we obtain the parameter set called $(a_\textrm{III},g_\textrm{III})$, that minimizes the distance
    $S(a_\textrm{I},g_\textrm{I},a_\textrm{III},g_\textrm{III})$. 
    From the matched pair $(a_\textrm{III},g_\textrm{III})$ we infer the mean free paths ${\ell}_{\textrm{tr}}^{\textrm{P}_N}(a_\textrm{III}, g_\textrm{III})$ and ${\ell}_{\textrm{abs}}^{\textrm{P}_N}(a_\textrm{III},g_\textrm{III})$ of that sample using the appropriate $P_N$ approximation. 
    This step is illustrated in Fig. \ref{fig:ExplanationComparison} where the blue grid point \textrm{III} in the $P_N$ grid represents ($a_\textrm{III}$,$g_\textrm{III}$) pair fitted to the blue grid point \textrm{I} in the  Monte Carlo grid. 
    
    \item 
    In the final step, the procedure of the previous step is repeated for all possible ($a_\textrm{I}$,$g_\textrm{I}$) and its {\it matched} pair ($a_\textrm{III},g_\textrm{III}$), and we calculate the relative errors of the interpretations of the analytic $P_N$ approximation:
    %
    \begin{eqnarray}{\label{eqn:ltrandlabs_ErrorCalc}}
            \Delta {\ell}_{\textrm{tr}}(a_\textrm{I},g_\textrm{I}) &\equiv \sqrt{\frac{\left[{\ell}_{\textrm{tr}}^{\textrm{MC}}(a_\textrm{I},g_\textrm{I}) -
            {\ell}_{\textrm{tr}}^{\textrm{P}_N}(a_\textrm{III},g_\textrm{III})\right]^2}{\left[{\ell}_{\textrm{tr}}^{\textrm{MC}}(a_\textrm{I},g_\textrm{I})\right]^2}} \times 100,\:\\
            \Delta {\ell}_{\textrm{abs}}(a_\textrm{I},g_\textrm{I}) &\equiv\sqrt{\frac{\left[{\ell}_{\textrm{abs}}^{\textrm{MC}}(a_\textrm{I},g_\textrm{I}) -
            {\ell}_{\textrm{abs}}^{\textrm{P}_N}(a_\textrm{III},g_\textrm{III})\right]^2}{\left[{\ell}_{\textrm{abs}}^{\textrm{MC}}(a_\textrm{I},g_\textrm{I})\right]^2}} \times 100.\:
    \end{eqnarray}
    These results are plotted in Fig. \ref{fig:ErrorMapltr&labs} for all $(a_\textrm{I},g_\textrm{I})$.
\end{enumerate}
%
\begin{figure*}[htb!]
    \includegraphics[width=1\textwidth]{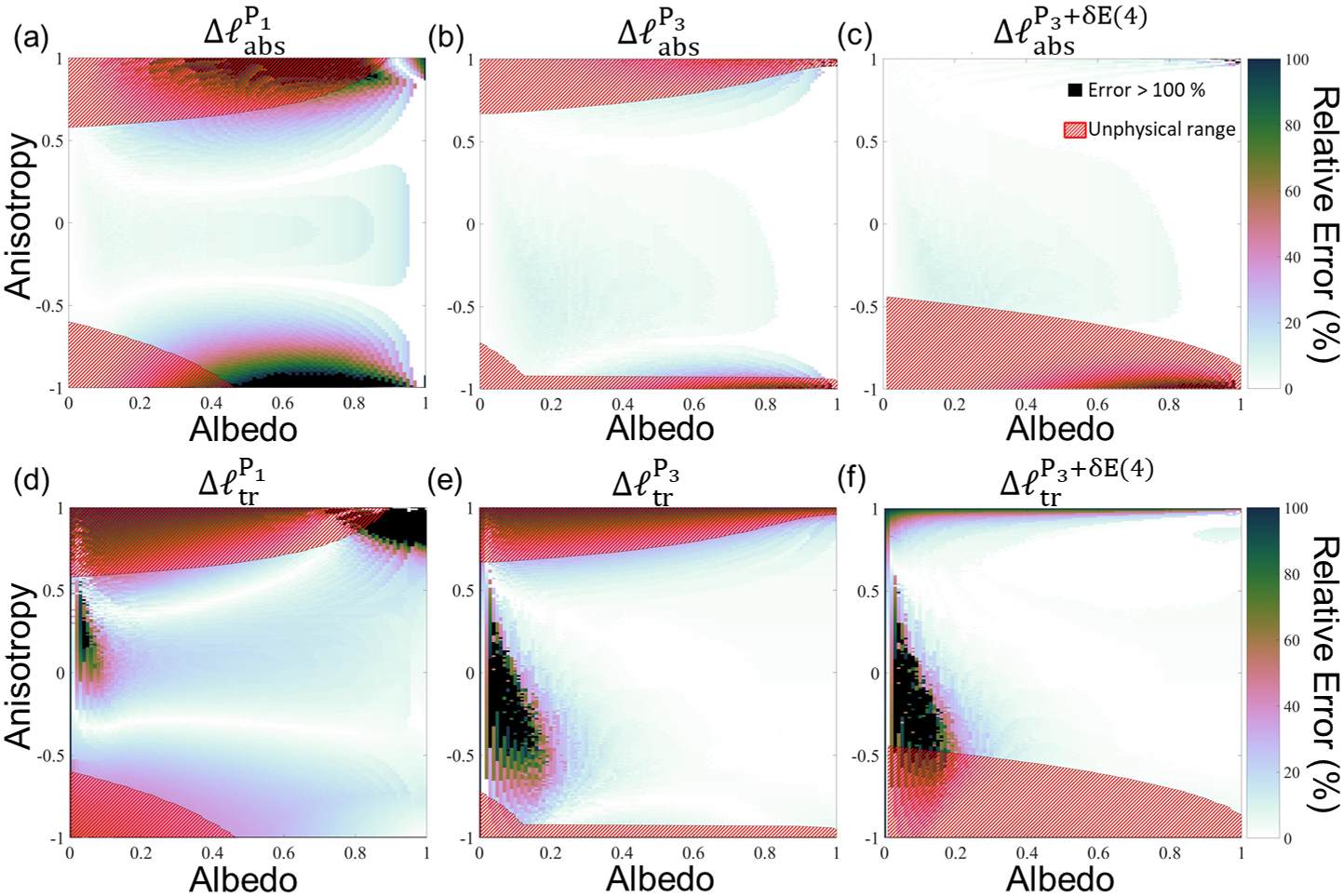}
    \caption{Relative error maps of absorption mean free path ${\ell}_{\textrm{abs}}$ and transport mean free path ${\ell}_{\textrm{tr}}$ for the $P_N$ approximations. 
    (a) and (d) for  $P_1$, (b) and (e) for $P_3$, and (c) and (f) for $P_{3}+\delta E(4)$. 
    Relative errors reported here are calculated by comparing these approximations to our Monte Carlo simulations, according to the coordinates found from the fitting process explained in the text and illustrated in Fig. \ref{fig:ExplanationComparison}. 
    Color map indicates the percentage of errors. 
    Values that are larger than $100\%$ are indicated with black markers. 
    The unphysical ranges of the approximations are added to the graphs as red hatched regions. 
    Optical thickness and refractive index contrast are specifically chosen as $b = 3$ and $\Delta n^2 = 0.245$ for these relative error maps.}
    \label{fig:ErrorMapltr&labs}
\end{figure*}

\subsection{Relative error maps inside and outside the physical ranges}
\label{subsection:ErrorMaps}
Fig. \ref{fig:ErrorMapTT&R} shows the relative error maps that present the deviations of the total transmission and the total reflection calculations of the analytical methods from the Monte Carlo simulations for realistic parameters $b = 3$ and $\Delta n^2 = 0.245$. 

Fig. \ref{fig:ErrorMapTT&R}(a) shows that the total transmission results of the $P_1$ approximation have up to $53\%$ relative error in the unphysical forward scattering region, and up to $86\%$ relative error in the backscattering regions. 
For the $P_1$ approximation, the relative errors observed in total transmission occur for more directional scattering, however, significant relative errors are also observed for the total reflection results in the isotropic regions ($g \approx 0$) reaching as high as $29\%$ for absorbing samples (see Fig. \ref{fig:ErrorMapTT&R}(d)).
For an isotropic sample the errors in total reflection decrease as the absorption decreases and are less than $5\%$ when the sample has no absorption ($a=1$).
Fig. \ref{fig:ErrorMapTT&R}(d) shows that the total reflection results of the $P_1$ approximation have large relative errors for most of the physical region, that can exceed $500\%$ for forward scattering samples with $g > 0.9$.
In the unphysical forward scattering region, the errors reach up to $10^3\%$.
These results show that the $P_1$ approximation is particularly bad at describing total reflection, even for photonic scattering slabs with weak absorption. 
This observation is in line with our observations that the unphysical behaviors are usually encountered near the incident (left) boundary of the slab, where the light enters the slab (see Figs. \ref{fig:ExamplesP1}(C,D)). 

Relative errors of total transmission of the $P_3$ approximation are given in Fig.~\ref{fig:ErrorMapTT&R}(b), which shows that it generally has less than $10\%$ error in its physical regions. 
For the unphysical forward scattering and backscattering cases, the errors are $25\%$ at most.
Similar to the $P_1$ approximation, in the extreme forward scattering unphysical regions ($g > 0.9$), the $P_3$ approximation has large relative errors that go up to $10^4\%$ in total reflection (see Fig.~\ref{fig:ErrorMapTT&R}(e)). 
For the backscattering unphysical region, the $P_3$ approximation has errors up to $21\%$ at most.
In the physical regions, the relative errors of total reflection calculations of the $P_3$ approximation go up to $100\%$ as the sample is in the vicinity of the unphysical forward scattering region. 
However, for the significant part of the physical region where there is weak anisotropy ($g<0.5$ \& $g>-0.5$), the relative errors are less than $10\%$.

Figs.~\ref{fig:ErrorMapTT&R}(c,f) show that the $P_3+\delta E(4)$ approximation has up to $73\%$ relative error in the total transmission for extremely backscattering samples ($g < -0.9$).
Even though the relative errors in total reflection results go up to $98\%$ for regions where $g>0.9$ and $a<0.1$, the remaining forward scattering range where $0.5<g<0.9$ is much more accurate than with the $P_1$ and $P_3$ approximations.

Fig.~\ref{fig:ErrorMapltr&labs} presents the deviations of the inferred mean free paths ${\ell}_{\textrm{tr}}^{\textrm{PN}}$ and ${\ell}_{\textrm{abs}}^{\textrm{PN}}$, predicted by analytic approximations using "measured" total transmission ($T^{\textrm{MC}}$) and total reflection ($R^{\textrm{MC}}$), from the "real" ${\ell}_{\textrm{tr}}^{\textrm{MC}}$ and ${\ell}_{\textrm{abs}}^{\textrm{MC}}$ transport parameters. 

Fig.~\ref{fig:ErrorMapltr&labs}(a) shows that the ${\ell}_{\textrm{abs}}^{P_1}$ has errors up to $211\%$ for the forward scattering unphysical range.
In addition, it is clear that the $P_1$ approximation also has large errors in the physical part of the forward scattering range.
In the unphysical backscattering range, the relative errors of ${\ell}_{\textrm{abs}}^{P_1}$ can be $100\%$, whereas, in the physical backscattering range, the errors reach $166\%$.
For the transport mean free path ${\ell}_{\textrm{tr}}^{P_1}$, Fig.~\ref{fig:ErrorMapltr&labs}(d) shows that the $P_1$ approximation has up to $56\%$ error for the backscattering range. 
In the forward scattering unphysical range, errors up to $100\%$ are observed.
In the physical, non-absorbing, and dominant forward scattering range, relative errors reach $10^7\%$.
Another high error region, for which errors up to $123\%$, is seen in Fig.~\ref{fig:ErrorMapltr&labs}(d), where there is strong absorption ($a<0.2$).

Fig.~\ref{fig:ErrorMapltr&labs}(b) shows that the $P_3$ approximation has up to $100\%$ relative error in the absorption mean free path ${\ell}_{\textrm{abs}}^{P_3}$ results for the strongly anisotropic samples $g < -0.9$ and $g > 0.9$. 
Moreover, the errors in the physical regions have fewer errors compared to the $P_1$ approximation.
The relative error map for ${\ell}_{\textrm{tr}}^{P_3}$ given in Fig.~\ref{fig:ErrorMapltr&labs}(e) shows relative errors up to $306\%$ in the strong absorption regions ($a<0.3$).
In the unphysical forward scattering range, the errors of ${\ell}_{\textrm{tr}}^{P_3}$ go up to $98\%$. 
In the rest of the grid, the $P_3$ approximation has significantly less relative errors than the $P_1$ approximation for extracting ${\ell}_{\textrm{tr}}$.

The relative error map of ${\ell}_{\textrm{abs}}^{P_3 + \delta E(4)}$ given in Fig.~\ref{fig:ErrorMapltr&labs}(c) shows that the $P_3 + \delta E(4)$ approximation has fewer regions with large errors compared to the $P_1$ and the $P_3$ approximations.
In the extremely anisotropic cases, however, the errors of ${\ell}_{\textrm{abs}}^{P_3 + \delta E(4)}$ reach $10^6\%$ for $g=0.98$, and $200\%$ for $g=-0.99$.
For ${\ell}_{\textrm{tr}}^{P_3 + \delta E(4)}$, Fig.~\ref{fig:ErrorMapltr&labs}(f) shows that fewer regions with large errors, especially in the forward scattering range, compared to the $P_1$ and the $P_3$ approximations.
In addition, Fig.~\ref{fig:ErrorMapltr&labs}(f) shows that the $P_3 + \delta E(4)$ approximation, similar to the $P_3$ approximation, has errors as large as $322\%$ in regions with strong absorption ($a<0.3$).

Fig.~\ref{fig:ErrorMapltr&labs} shows that the $P_1$ approximation has more regions in the grid with large relative errors, as expected. 
As the absorption increases, the $P_3$ and the $P_3+\delta E(4)$ approximations predict results with relative errors more than $100\%$, however, both are more accurate for the rest of the grid, with the $P_3+\delta E(4)$ approximation performing better for a wider forward scattering range.

With prior knowledge of parameters ($a$,$g$,$b$,$\Delta n^2$), one can verify if a sample is in the unphysical ranges and use the relative error maps provided in this work to see if any of the reported $P_N$ approximations are appropriate for their sample.
Needless to say, if ($a$,$g$,$b$,$\Delta n^2$) are already known, one could get the transport parameters $\ell_{\textrm{tr}}$ and $\ell_{\textrm{abs}}$ without the use of $P_N$ approximation, given that the density of scatterers in their sample fulfills the independent scattering approximation~\cite{Lagendijk1996PR}. 
Nevertheless, it is necessary to use either analytical approximations or numerical methods to infer the position-dependent energy density $u(z)$ and the diffuse flux $F(z)$ inside the sample.
Hence, for these investigations, the unphysical ranges and relative error maps of the analytical methods are crucial.

\subsection{Practical cases}
\label{subsection:Practical}
A practical example is a slab of human dermis in air. 
The optical constants of this slab are~\cite{Jacques1987LLS}: $a=0.99$, $g=0.81$, and $\Delta n^2 = 0.245$ (assuming $n_\textrm{slab} = 1.4$). 
If the slab has an optical thickness $b=3$, Figs.~\ref{fig:ErrorMapTT&R} and~\ref{fig:ErrorMapltr&labs} serve to choose the analytical approximation to infer the position-dependent energy density and diffuse flux inside the tissue.
In this case, the $P_3 + \delta E(4)$ approximation would be the best choice, as it has less than $1\%$ error for extracting ${\ell}_\textrm{tr}$ and ${\ell}_\textrm{abs}$.
This decision is in line with Star~\cite{Star1989BookChp}, who investigated a thicker ($b=9.5$) slab of human dermis and found the $P_3 + \delta E(4)$ approximation to be more accurate than the $P_3$ approximation.

Other examples stem from the solid state lighting industry, where understanding the light transport in white LEDs is an important issue.
A typical white LED contains a phosphor layer that absorbs incoming blue light and re-emits light in different wavelengths to achieve white output.
In reference~\cite{Meretska2017OE}, polymer slabs containing phosphor (YAG:Ce\textsuperscript{3+}) scatterers are studied, and their optical constants are found using Monte Carlo simulations and the $P_3$ approximation.
The refractive index of the polymer matrix is $n_\textrm{slab}=1.4$, and the surrounding medium is air, which gives $\Delta n^2 = 0.245$. 
The optical constants for such a slab are derived to be~\cite{Meretska2017OE}: $a = 0.89$, $g = 0.72$ at a wavelength $\lambda = 460\text{nm}$ where the phosphors are absorbing, and $a = 1$, $g = 0.82$ at $\lambda = 600\text{nm}$ where the phosphors are non-absorbing.
If we consider a slab with optical thickness $b=3$ and look at the Figs.~\ref{fig:ErrorMapTT&R} and~\ref{fig:ErrorMapltr&labs}, we see that the $P_3 + \delta E(4)$ approximation would be the best one to use for both cases since it has less than $1\%$ error for extracting ${\ell}_\textrm{tr}$ and ${\ell}_\textrm{abs}$, whereas the $P_3$ approximation has $\Delta {\ell}_\textrm{tr}^{P_3} = 7\%$ for absorbing, and $\Delta {\ell}_\textrm{tr}^{P_3} = 10\%$ for non-absorbing wavelengths.

Earlier on, Jacques \textit{et al.} have experimentally and numerically investigated light distribution in various phosphor plates~\cite{Krasnoshchoka2020OE}. 
The optical constants of a slab containing composite ceramic Ce:YAG are reported as: $a = 0.71$, $g = 0.75$ at a wavelength $\lambda = 450\text{nm}$ where the phosphors are absorbing~\cite{Krasnoshchoka2020OE}. 
Assuming the index contrast $\Delta n^2 = 0.245$, and the optical thickness $b=3$, Figs.~\ref{fig:ErrorMapTT&R} and~\ref{fig:ErrorMapltr&labs} shows less than $1\%$ error for $\Delta {\ell}_\textrm{tr}^{P_3 + \delta E(4)}$ and $\Delta {\ell}_\textrm{abs}^{P_3 + \delta E(4)}$. 
The lengths extracted using the $P_3$ approximation have $\Delta {\ell}_\textrm{tr}^{P_3} = 16.15\%$ and $\Delta {\ell}_\textrm{abs}^{P_3} = 7.41\%$, and with the $P_1$ approximation have $\Delta {\ell}_\textrm{tr}^{P_1} = 0.4 \%$ and $\Delta {\ell}_\textrm{abs}^{P_1} = 52.63\%$.
A more absorbing slab containing Ce:LuAG and Eu:nitride has reported optical constants $a=0.25$ and $g=0.75$ at a wavelength $\lambda = 450\text{nm}$~\cite{Krasnoshchoka2020OE}. 
With $\Delta n^2 = 0.245$ and $b=3$, Figs.~\ref{fig:ErrorMapTT&R} and~\ref{fig:ErrorMapltr&labs} shows that the $P_3 + \delta E(4)$ approximation yields $\ell_\textrm{tr}$ and $\ell_\textrm{abs}$ with less than $1\%$ relative error, whereas with the $P_1$ approximation have $\Delta {\ell}_\textrm{tr}^{P_1} = 62.26 \%$ and $\Delta {\ell}_\textrm{abs}^{P_1} = 38.89\%$, and the $P_3$ approximation have $\Delta {\ell}_\textrm{tr}^{P_3} = 40.13 \%$ and $\Delta {\ell}_\textrm{abs}^{P_3} = 5.63 \%$.
For both slabs, the $P_3 + \delta E(4)$ is the best analytical approximation out of all three, as it gives the least relative errors, in agreement with our discussions above.

\section{Conclusions}
We have studied the unphysical ranges of the $P_1$, $P_3$, and $P_3+\delta E(4)$ approximations to the radiative transfer equation, for the complete ($a$,$g$,$b$,$\Delta n^2$) parameter space. 
The unphysical parameter ranges are characterized by unphysical negative energy densities and fluxes of the wrong sign. 
These ranges are crucial when the position-dependent energy density inside the photonic scattering slab is being investigated. 

Typically researchers want to extract the transport parameters ${\ell}_{\textrm{tr}}$ and ${\ell}_{\textrm{abs}}$ from total transmission and reflection experiments. 
We have calculated the relative errors in the transport parameters for all possible albedo and anisotropy and for realistically chosen optical thickness and refractive index contrast, by comparing the analytical results with Monte Carlo simulations.

In the unphysical ranges the relative errors are as large as $10^4\%$, but also in the physical ranges the errors can be substantial. 
We emphasize that the relative error maps provided here are for slabs with widely studied realistic optical thickness and refractive index contrast. 
Maps for any other kind of samples with their specific parameters require characterization using the methodology explained here.

We conclude that the $P_1$ approximation is not viable to extract either the transport parameters or the position dependent energy density, unless the scattering of the sample is purely isotropic and elastic. 
The $P_3$ and $P_3 + \delta E(4)$ approximations are safer to use than the $P_1$ approximation, unless there is strong absorption ($a < 0.3$) or extreme anisotropy ($g > 0.9$ and $g < -0.9$). 
The approximations should not be used if the samples are in the unphysical parameter range, even though the relative errors are low in some parts of these unphysical ranges. 
Especially, the $P_3 + \delta E(4)$ approximation is suited for enhancing the accuracy in the forward direction ($g > 0$) and should not be used in the backscattering range ($g < 0$). 
Our results provide a guideline for the applicability of the $P_1$, $P_3$, and $P_3+\delta E(4)$ approximations to the radiative transfer equation, to interpret experiments on light transport in photonic scattering slabs.

\section{Acknowledgements}
We thank Marek Kozoň, Lars Corbijn van Willenswaard, Minh Duy Truong and Matthias Schlottbom for useful discussions. 
We thank Maryna Meretska and Ravitej Uppu for early contributions to this project. 
We thank Wilbert IJzerman and Gilles Vissenberg (Signify), Niels van der Veen, Helmut Bechtel, and Toni Lopez (Lumileds) for practical advice.
This work was supported by the NWO-TTW program P15-36 ”Free-form scattering optics” (FFSO) and the MESA+ Institute section Applied Nanophotonics (ANP).

The data used for this publication are available via the open-access repository Zenodo database that is developed under the European OpenAIRE program and operated by CERN~\cite{ZenodoDoi}. 

\appendix
 
\section{\texorpdfstring{$P_N$}{TEXT} approximation}\label{appendix:PNapproximation}

We consider a slab with well-dispersed spherical scatterers as the photonic scattering medium (see Fig.~\ref{fig:Illustration}). 
The RTE is then
%
\begin{equation}{\label{eqn:RTE_Slab}}
\begin{split}
    \mu\frac{dI_{\textrm{diff}}(z,\mu)}{dz} &=  -\rho\sigma_\textrm{t} I_{\textrm{diff}}(z,\mu) \\ 
    &+ \rho\sigma_\textrm{t} a (2\pi) \int_{-1}^{1} p(\mu,\mu')I_{\textrm{diff}}(z,\mu')d\mu' \\
    &+ \epsilon(z,\mu).
    \end{split}
\end{equation}
%
Eq. (\ref{eqn:RTE_Slab}) describes the change in diffuse specific intensity ($I_{\textrm{diff}}$) with position, in the direction $\mu$ ($\mu = \cos{\theta}$; $\theta$ being the angle between the direction of light and $\hat{z}$, a.k.a. the scattering angle). 
Here, $\rho$ is the density of scatterers, $\sigma_\textrm{t}$ is the extinction cross-section, and $a$ is the albedo. 
The first term on the right-hand side describes the decrease in $I_{\textrm{diff}}$ due to scattering and absorption events. 
The second term describes the increase in $I_{\textrm{diff}}$ due to light coming from other directions and scattering into the direction we are looking at. 
The phase function $p(\mu,\mu')$ describes the probability of light coming from $\mu'$ to scatter into $\mu$ direction. 
The third is the external source term, for which we use plane waves.

In the $P_N$ approximation, $I_{\textrm{diff}}$ is expressed as
%
\begin{equation}{\label{eqn:IdiffExpanded}}
    I_{\textrm{diff}}(z,\mu) = \sum_{l = 0}^{\infty} \frac{2l+1}{2} \psi_l(z) P_l(\mu),
\end{equation}
%
where, $P_l(\mu)$ is a Legendre polynomial and $\psi_l(z)$ is
%
\begin{equation}{\label{eqn:psiMain}}
    \psi_l(z) = \int_{-1}^{1} I_{\textrm{diff}}(z,\mu) P_l(\mu) d\mu.
\end{equation}
%
The expansion in Eq. (\ref{eqn:IdiffExpanded}) is obtained by expanding $I_{\textrm{diff}}$ in terms of spherical harmonics, as in Eq. (\ref{eqn:specIntExp}), and omitting the azimuthal term in the expansion. 
The azimuthal term vanishes after integration over $4\pi$ due to the symmetries introduced by the well-dispersed spherical scatterers in the slab. 
Similarly, the source term $\epsilon(z,\mu)$ is expanded as
%
\begin{equation}{\label{eqn:SourceExpanded}}
    \epsilon(z,\mu) = \sum_{l = 0}^{\infty} \frac{2l+1}{2} s_l(z) P_l(\mu),
\end{equation}
%
where $s_l(z)$ is the $l^{th}$ moment of the source function. For plane waves as the source, 
%
\begin{equation}
s_l(z) = F_0 \rho \sigma_\textrm{t} a w_l e^{-\rho\sigma_\textrm{t} z},
\end{equation}
%
where $F_0$ is the incident flux.

The phase function of the system only depends on the angle between the incoming $\hat{s}'$ and outgoing $\hat{s}$ directions. 
Thus, we can expand the phase function as
%
\begin{equation}{\label{eqn:phasefunc1}}
    p(\hat{s},\hat{s'}) = \sum_{l = 0}^{\infty} \frac{2l+1}{4\pi} w_l P_l(\hat{s}\cdot\hat{s'}).
\end{equation}
%
Using the addition theorem for spherical harmonics \cite{Arfken2005Book}, we get
%
\begin{equation}{\label{eqn:AdditionTheorem_P_l(alpha)}}
\begin{split}
    &P_l(\hat{s}\cdot\hat{s'}) = P_l(\mu)P_l(\mu') \:+ \\
    &\left[\sum_{m = -l}^{l} \frac{(l-m)!}{(l+m)!} P_l(\mu)^m P_l(\mu')^m cos\left[m(\phi'-\phi)\right]\right],
\end{split}
\end{equation}
%
where $\phi$ and $\phi'$ are azimuthal angles. 
Due to symmetries in the slab mentioned above, $m = 0$ and the second term on the right-hand side vanishes after integration over $4\pi$. 
Thus, we rewrite (\ref{eqn:phasefunc1}) as
%
\begin{equation}{\label{eqn:phaseExpanded}}
    p(\hat{s},\hat{s'}) = p(\mu,\mu') = \sum_{l = 0}^{\infty} \frac{2l+1}{4\pi} w_l P_l(\mu) P_l(\mu').
\end{equation}
%
By using the well-known Henyey-Greenstein phase function \cite{Henyey-Greenstein1941APJ} we get
%
\begin{equation}
    w_l = g^l, 
\end{equation}
%
where $g$ is the anisotropy. 

Substituting equations (\ref{eqn:IdiffExpanded}), (\ref{eqn:SourceExpanded}) and (\ref{eqn:phaseExpanded}) into equation (\ref{eqn:RTE_Slab}), we arrive at the final expression;
%
\begin{equation}{\label{eqn:PnFinalExpression}}
\begin{split}
     (l+1)\frac{d\psi_{l+1}}{dz} &+ l\frac{d\psi_{l-1}}{dz} \\
     &+ (2l+1)(1-a g^l)\rho\sigma_\textrm{t}\psi_l(z) \\
     &= (2l+1) F_0 \rho \sigma_\textrm{t} a g^l e^{-\rho\sigma_\textrm{t} z} ,
\end{split}
\end{equation}
%
which is an infinite set of coupled differential equations that describe light propagation in the scattering medium.

While we stated in Section~\ref{section:Introduction} that the solution is fully determined if the four-dimensional parameter set $(a,g,b,\Delta n^2)$ is known, but inspection of equation~(\ref{eqn:PnFinalExpression}) does not immediately make this claim obvious.
The parameters $a$ and $g$ are explicitly present in equation~(\ref{eqn:PnFinalExpression}). 
To solve a differential equation boundary conditions are necessary and that is where $b$ and $\Delta n^2$ show up. 
However, it seems that the differential equations also depend on the independent parameter $\rho\sigma_\textrm{t}$. 
But the parameter $\rho\sigma_\textrm{t}$, which is the inverse of the extinction length, can be scaled out of the equation, if $z$ is rescaled as$z \rho\sigma_\textrm{t}$. 

To solve this system of differential equations, we limit the expansion to order $N$. 
An odd positive integer N is chosen with
%
\begin{equation}
\begin{split}
    \psi_{-1}(z) &= 0, \\
    N<l<\infty &\implies \psi_l(z) = 0.
\end{split}    
\end{equation}
%
Depending on $N$, we get a set of $N+1$ differential equations to solve, as $l$ only takes values $l = 0,1,...,N$. 
This set is solved with the same method used in Refs.~\cite{Star1989BookChp, Meretska2019ACSPhot} to find $\psi_0(z)$ and $\psi_1(z)$, to obtain the average intensity
%
\begin{equation}
    U(z) = \frac{1}{4\pi} \psi_0(z),
\end{equation}
%
and the diffuse flux
%
\begin{equation}
    F(z) = \psi_1(z).
\end{equation}
%

%
\begin{table*}[htb!]
    \begin{adjustbox}{width=1\textwidth}
            \begin{tabular}{lcccccrclcrclcrcl}
            \hline
            \hline
            \textbf{Method}&  \textbf{a} & \textbf{g} & \textbf{b} & $\boldsymbol{n_{\textrm{slab}}}$ & & \multicolumn{3}{c}{\textbf{Total Transmission}} & & \multicolumn{3}{c}{\textbf{Reflection}} & & \multicolumn{3}{c}{\textbf{Absorption}}     \\
            \hline
            Lambert-Beer$^{a}$ & \multirow{2}{*}{$0$} & \multirow{2}{*}{$0.75$} & \multirow{2}{*}{$3$} & \multirow{2}{*}{$1.4$} & & $4.8404\% $ & & & & &-& & & &-&  \\
            Monte Carlo &  &  &  &  & & $4.8404\% $ &$\pm$& $0.0059\% $ & & $0.00670\% $ &$\pm$& $0.00003\% $ & & $95.155\% $ &$\pm$& $0.006\% $ \\
            \hline
            Intensity Fabry-Pérot & \multirow{2}{*}{$1$} & \multirow{2}{*}{$1$} & \multirow{2}{*}{$3$} & \multirow{2}{*}{$1.4$} & & $97.2973\%$ & & & & $2.7027\% $ & & & & & - & \\
            Monte Carlo &  &  &  &  & & $97.2973\% $ &$\pm$& $0.0006\%$ & & $2.70270\%$ &$\pm$& $0.00005\%$ & & $0.00\%$ & & \\
            \hline
            Diffusion ($P_1$ approx.) & \multirow{2}{*}{$1$} & \multirow{2}{*}{$0$} & \multirow{2}{*}{$100$} & \multirow{2}{*}{$1.5$} & & $3.24\%$ & & & & $96.70\% $ & & & & $0.00\% $ & &  \\
            Monte Carlo &  &  &  &  & & $3.231\% $ &$\pm$& $0.005\% $ & & $96.767\% $ &$\pm$& $0.005\% $ & & $0.00\% $ & & \\
            \hline
            van de Hulst \cite{vanDeHulst1980Book} & \multirow{2}{*}{$0.9$} & \multirow{2}{*}{$0.75$} & \multirow{2}{*}{$2$} & \multirow{2}{*}{$1$} & & $66.096\% $ & & & & $9.739\% $ & & & & & - & \\
            Monte Carlo &  &  &  &  & & $66.094\% $ &$\pm$& $0.012\% $ & & $9.740\% $ &$\pm$& $0.008\% $ & & $24.166\% $ &$\pm$& $0.008\% $ \\
            \hline
            \hline
            \end{tabular}
    \end{adjustbox}
\caption{Verification of our Monte Carlo simulations. Absorbing ($a=0$), non-absorbing ($a=1$), isotropic ($g=0$), and forward scattering ($g \geq 0.75$) cases are compared. Reported errors are standard deviations of the mean, calculated from ten runs of each 10 million photons. Only for the $b = 100$ case, the error is calculated from ten runs of each 1 million photons. \textbf{(a)} Lambert-Beer calculation is done for the $97.22\%$ fraction of the incident light since the initial reflection from the left boundary is not included in our Monte Carlo simulations.}
\label{tab:VerificationTable}
\end{table*}
%

\section{\texorpdfstring{$P_N + \delta E(N+1)$}{TEXT} approximation}\label{appendix:PNdE(4)}

When the order $N$ of the approximation is chosen, the higher-order terms of the expansions are truncated. 
Therefore, the implementations of higher-order approximations are expected to be more accurate, especially when it comes to photonic scattering slab with anisotropic scattering. 
$P_N + \delta E(N+1)$ approximation adds a delta function to the Henyey-Greenstein phase function to compensate for the loss in extreme forward scattering due to truncated higher-order terms\cite{Star1988PMB,Meador1979AO}. 
As the RTE and its moments keep their mathematical form, only the optical constants need to be modified as\cite{Star1989BookChp}:
%
\begin{equation}{\label{eqn:ModifiedConstantsForDeltaE}}
\begin{split}
    \sigma_\textrm{s}' = (1-f)\sigma_\textrm{s} \:\: &; \:\: (g')^n = \frac{g^n - f}{1-f} \: ;\\
    f = g^{N+1} \:\: &; \:\: a' = \frac{\sigma_\textrm{s}'}{\sigma_\textrm{a} + \sigma_\textrm{s}'}.
\end{split}
\end{equation}
%
Replacing the original optical constants with the ones in equation (\ref{eqn:ModifiedConstantsForDeltaE}) and taking the order $N = 3$ in the method explained in Appendix \ref{appendix:PNapproximation}, we get the solution for $P_3 + \delta E(4)$ approximation to RTE.

\section{Monte Carlo simulations}\label{appendix:MonteCarlo}

The Monte Carlo simulations used in this work follow the same principles as the work of Prahl et al.~\cite{Prahl1989BookChp} and of Jacques~\cite{Jacques2011BookChp}. 
In brief, one photon is repeatedly launched inside the photonic scattering slab, with a "photon weight" of 1 and initial direction $\hat{z}$. 
The photon moves a step length based on the probability of photon travel before getting absorbed or scattered inside the slab. 
After that step, a fraction of photon weight (determined by the albedo) is deposited at the local bin in that location, and the remaining weight is propagating into new direction, determined by using the Henyey-Greenstein phase function for the specified anisotropy. 
The propagation is continued by generating new steps until the photon escapes a boundary or its weight is below a threshold.

The main difference of our simulation code with those of Prahl \textit{et al.}~\cite{Prahl1989BookChp} and Jacques~\cite{Jacques2011BookChp} is the incorporation of multiple internal reflection of photons from the boundaries of the slab (see Fig. \ref{fig:Illustration}). 
If an extremely long step length is generated, a fraction of the photon weight could end up internally reflecting multiple times. 
This situation is considered in our simulations by the addition of intensity Fabry-Perot calculations. 

The Monte Carlo simulations are verified by comparing their results with exact calculations (and diffusion approximation for a slab with very long thickness $b = 100$). 
The comparisons are reported in Table \ref{tab:VerificationTable}. 
The Lambert-Beer law used in Table \ref{tab:VerificationTable} is
\begin{equation}
    I_\textrm{b} = I_\textrm{0} e^{-\rho\sigma_\textrm{t} L} \equiv I_\textrm{0} e^{-b},
\end{equation}
where $I_\textrm{b}$ is the transmitted ballistic flux.
We compare this with the case where $a=0$, so the total transmission only has the ballistic component as the diffuse part is completely absorbed. 
For the purely forward scattering and non-absorbing case ($a=1$; $g=1$) we compared our Monte Carlo simulations to the Intensity Fabry-Pérot calculations, for which only the internal reflections from the boundaries of the slab determines the total transmission and reflection of the slab. 
In addition, an anisotropic case with weak absorption ($g=0.75$; $a=0.9$) is compared with the calculations reported by H.C. van de Hulst in Table 35 of \cite{vanDeHulst1980Book}. 
It is clear from Table \ref{tab:VerificationTable} that our Monte Carlo simulations are in good agreement with all of the compared methods, which indicates the high accuracy of our simulations.

\bibliographystyle{apsrev4-1}
\bibliography{_RTE_breakdown.bib}

\end{document}